      [1999/12/01 v1.4c Il Nuovo Cimento]
\documentclass{cimento}
\usepackage{url}
\usepackage{relsize}
\def\babar{\mbox{\slshape B\kern-0.1em{\smaller A}\kern-0.1em
    B\kern-0.1em{\smaller A\kern-0.2em R}}}

\newcommand{\power}[1]{\times 10^{#1}}
\newcommand{\nn}{\nonumber}
\newcommand{\bary}{\begin{array}}
\newcommand{\eary}{\end{array}}

\newcommand{\lsim}{\mbox{\raisebox{-0.3ex}{\footnotesize $\:\stackrel{<}{\sim}\:$}} }
\newcommand{\sinW}{\sin^2 \Theta_W}
\newcommand{\Gmu}{G_{\mu}}
\newcommand{\dro}{\Delta \rho}
\newcommand{\cosi}{\cos^2 \Theta_i}
\newcommand{\sini}{\sin^2 \Theta_i}
\newcommand{\sing}{\sin^2 \Theta_g}
\newcommand{\gv}{\mbox{GeV}}

\newcommand{\mv}{\mbox{MeV}}

\newcommand{\ha}{\frac{1}{2}}
\newcommand{\mbo}[1]{$ #1 $}

\newcommand{\cs}{\;,\;\;}
\newcommand{\epo}{\;. }
\newcommand{\varv}{v}
\newcommand{\dal}{\Delta \alpha}

\newcommand{\seffl}{\sin^2\theta ^{\mathrm{lep}}_{\mathrm{eff}}}
\newcommand{\sigh}{$\sigma(\epm \to hadrons)$}
\newcommand{\alqed}{$\alpha_\mathrm{QED}$}

\newcommand{\sinf}{\sin^2 \Theta_f}
\newcommand{\al}{\alpha}

\newcommand{\Repa}{\mbox{Re} \:}

\newcommand{\be}{\begin{equation}}
\newcommand{\ee}{\end{equation}}
\newcommand{\ba}{\begin{eqnarray}}
\newcommand{\ea}{\end{eqnarray}}

\newcommand{\crn}{\nn \\}
\newcommand{\epm}{e^+e^-}

\def\mz{M_Z^2}
\def\MZ{M_Z}

\def\das{\Delta\alpha(s)}

\def\dah{\Delta\alpha^{(5)}_{\rm had}}
\def\dahs{\Delta\alpha^{(5)}_{\rm had}(s)}

\def\dahz{\Delta\alpha^{(5)}_{\rm had}(\MZ^2)}
\def\dah0{\Delta\alpha^{(5)}_{\rm had}(-s_0)}

\usepackage{graphicx}  

\title{Electroweak effective couplings for future precision experiments}
\author{F. Jegerlehner\from{ins:hub}\from{ins:zeu}
}
\instlist{\inst{ins:hub} 
Humboldt-Universit\"at zu Berlin, Institut f\"ur Physik,
       Newtonstrasse 15, D-12489 Berlin, Germany
  \inst{ins:zeu} Deutsches Elektronen-Synchrotron (DESY), 
Platanenallee 6, D-15738 Zeuthen, Germany}
\PACSes{\PACSit{12.15.-y,13.40.-f,12.15.Lk,13.40.Ks}{}
}

\DeclareMathSymbol{\varPhi}{\mathalpha}{operators}{"08}
\DeclareMathSymbol{\varOmega}{\mathalpha}{operators}{"0A}

\begin{document}
\renewcommand{\thefootnote}{\fnsymbol{footnote}}
\setlength{\baselineskip}{0.52cm}
\thispagestyle{empty}
\begin{flushright}
HU-EP-11/33 \\
DESY 11--117 \\
July 2011\\
\end{flushright}

\setcounter{page}{0}

\mbox{}
\vspace*{\fill}
\begin{center}
{\Large\bf 
Electroweak effective couplings for future precision experiments} \\

\vspace{5em}
\large
F. Jegerlehner\footnote[4]{\noindent Invited talk at 
LC10 Workshop ``New Physics: complementarities between direct and indirect searches'',
November 30th - December 3rd, 2010, INFN Frascati National
Laboratories, Frascati, Italy. 
This work
was supported in part by the European Commission's TARI program under contract
RII3-CT-2004-506078.} \\
\vspace{5em}
\normalsize
{\it Humboldt-Universit\"at zu Berlin, Institut f\"ur Physik,
       Newtonstrasse 15, D-12489 Berlin, Germany}\\
{\it Deutsches Elektronen-Synchrotron (DESY), 
Platanenallee 6, D-15738 Zeuthen, Germany}
\end{center}
\vspace*{\fill}
\newpage

\maketitle


\begin{abstract}
The leading hadronic effects in electroweak theory derive from vacuum
polarization which are non-perturbative hadronic contributions to the
running of the gauge couplings, the electromagnetic
$\alpha_\mathrm{em}(s)$ and the $SU(2)_L$ coupling $\alpha_2(s)$. I will
report on my recent package {\tt alphaQED}~\cite{alphaQED}, which besides the effective fine
structure constant $\alpha_{\rm em}(s)$ also allows for a fairly precise
calculation of the $SU(2)_L$ gauge coupling $\alpha_2(s)$.  I will
briefly review the role, future requirements and possibilities.
Applied together with the {\tt Rhad} package by Harlander and
Steinhauser~\cite{Harlander:2002ur}, the package allows to calculate
all SM running couplings as well as running $\sin^2\Theta$ versions
with state-of-the-art accuracy.
\end{abstract}

\section{Introduction}
Precise Standard Model (SM) predictions require to determine the 
{ $U(1)_Y\otimes
SU(2)_L\otimes SU(3)_c$} 
SM gauge couplings {$\al_\mathrm{em}$,
$\alpha_2$ and $\al_s\equiv\alpha_3$ (QCD)}  
as accurately as possible. Obviously, the predictability of theory is
limited by the precision of its input parameters. 
This in particular requires to fight precision limitations due to
non-perturbative hadronic contributions. Precise predictions
confronting precise measurements are the basis for all SM
precision tests, which allow us to unravel new physics from
discrepancies between theory and experiment. An important test case,
which requires as precise as possible running couplings, is the
quest of gauge coupling unification in grand unified extensions of the
SM.

Key input parameter for ILC physics currently are known to precision:\\
\be \bary{cccccccccccc}
\frac{\delta \alpha}{\alpha} &\sim& 3.6 &\times& 10^{-9}&~~~~~&
\frac{\delta \alpha(M_Z)}{\alpha(M_Z)} &\sim& {1.6 \div 6.8}
&{\times}&{10^{-4}} &\\
\frac{\delta G_\mu}{G_\mu} &\sim& 8.6 &\times& 10^{-6}&&
\frac{\delta M_Z}{M_Z} &\sim& 2.4 &\times& 10^{-5} & \epo
\eary
\ee
We observe that the accuracy of $\alpha(M_Z)$ is roughly one order of
magnitude worse than that of the next best $M_Z$! The loss in
precision caused by non-perturbative strong interaction effects is
$10^{5}$ between the classical low energy $\alpha$ and
$\alpha(M_Z)$. The requirement for ILC precision physics is
\be
\frac{\delta \alpha(M_Z)}{\alpha(M_Z)} \sim 5\power{-5}\epo
\ee 
A prominent example where theory may be obscured by lack of precision
in the effective $\alpha$ is the indirect Higgs mass bound obtained
from the precise measurement of $\seffl$. The required improvement
could be achieved by dedicated efforts in cross-section measurements
in the energy range from 1.2 to 3.2 GeV, and be adopting the Adler function
controlled split in parts evaluated from data (from experiments or from lattice QCD
simulations) and parts which can be calculated reliably in
perturbative QCD (pQCD):
\ba
\Delta\alpha^{(5)}_{\rm had}(M_Z^2)&=&\Delta\alpha^{(5)}_{\rm had}(-s_0)^{\mathrm{data}}+
\left[\Delta\alpha^{(5)}_{\rm had}(-M_Z^2) -\Delta\alpha^{(5)}_{\rm
had}(-s_0)\right]^{\mathrm{pQCD}} \crn &&
\left[\Delta\alpha^{(5)}_{\rm had}(M_Z^2) -\Delta\alpha^{(5)}_{\rm
had}(-M_Z^2)\right]^{\mathrm{pQCD}}\cs
\label{eq:alphaAdler}
\ea
where $s_0$ can be optimized by adopting the Adler function as a
monitor for the range of validity of
pQCD~\cite{Eidelman:1998vc,Jegerlehner:2008rs}. In
the following we will present a description of the package {\tt
alphaQED} which allows state-of-the-art calculations of the SM running
couplings, optionally, with their imaginary parts. Some emphasis is
put on the not so straight forward determination of the running
$SU(2)_L$ coupling $\alpha_2(s)$, which is important for the
calculation of variants of the weak mixing parameter $\sin^2
\Theta_W(s)$, an interesting quasi-observable and monitor of new
physics particularly at ILC energy scales.

\section{Effective running coupling \alqed}
\begin{figure}
\centering
\includegraphics[height=9cm]{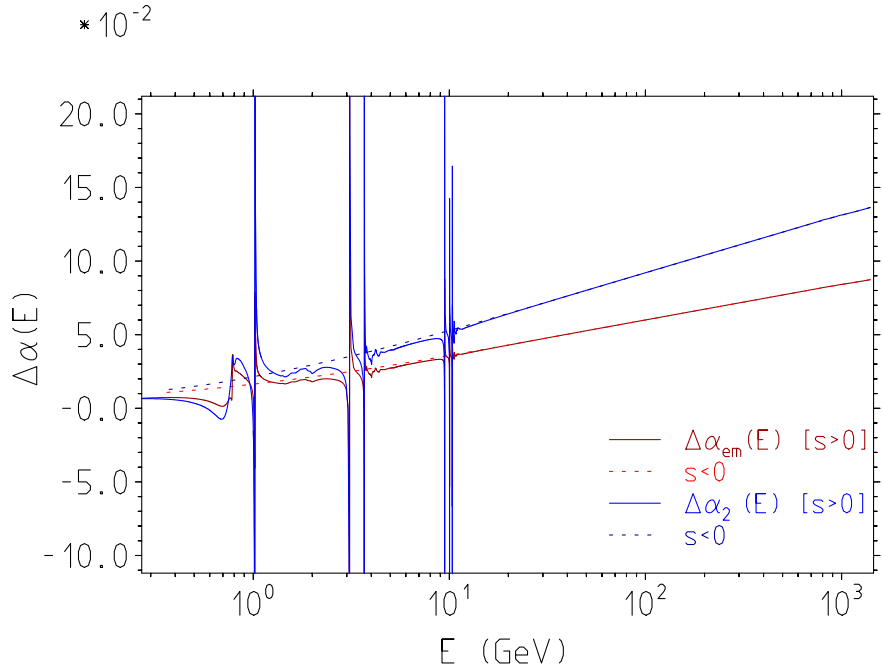}

\caption{$\Delta \alpha_{\rm em}(E)$ and $\Delta \alpha_{2}(E)$ as functions of energy 
$E$ in the time-like and space-like domain.
The smooth space-like correction (dashed line) agrees rather well
with the non-resonant ``background'' above the $\phi$-resonance (kind
of duality). In resonance regions as expected ``agreement'' is
observed in the mean, with huge local deviations.}
\label{fig:alphaQEDdual} 
\end{figure}

The effective fine-structure ``constant'' $\alpha(E)$
depends on the energy scale because of charge screening by vacuum polarization:
\ba
\das &=&-e^2\, \left[\Repa \Pi'^{\gamma \gamma}(s)
-\Pi'^{\gamma \gamma}(0)\right]
\label{eq:defdal}
\ea
which exhibit the leading hadronic non-perturbative part
$\Delta^{(5)}_{\mathrm{had}} \alpha$.  $\Pi(s)=\Pi(0)+s\,\Pi'(s)$
denotes the transversal current correlator, for the electromagnetic
current $\Pi(0)=0$. While electroweak effects (leptons etc.) are
calculable in perturbation theory, the calculation of the strong
interaction effects (hadrons/quarks etc.) by perturbative QCD 
fails. Fortunately, dispersion relations and the optical theorem allow
us to perform rather accurate evaluations in terms of experimental
$\epm$--data encoded in
\ba
R_\gamma(s) \equiv \frac{\sigma(e^+e^- \rightarrow \gamma^*
\rightarrow {\rm hadrons})}{ \sigma(e^+e^- \rightarrow \gamma^* \rightarrow
\mu^+ \mu^-)} \epo
\ea
For the electromagnetic running coupling the dispersion integral reads
\ba
\dahs &=& - \frac{\alpha s}{3\pi}\;\bigg(\;\;\;
{\rm \footnotesize P}\!\!\!\!\!\!\!\!  \int\limits_{4m_\pi^2}^{E^2_{\rm cut}} ds'
\frac{{R^{\mathrm{data}}_\gamma(s')}}{s'(s'-s)}
+ {\rm \footnotesize P}\!\!\!\!\!\!\!\!
\int\limits_{E^2_{\rm cut}}^\infty ds'
\frac{{ R^{\mathrm{pQCD}}_\gamma(s')}}{s'(s'-s)}\,\,
\bigg)
\label{eq:DI}
\ea
The high energy tail is neatly calculable perturbatively by the virtue
of asymptotic freedom of QCD.  Errors of data imply theoretical
uncertainties. Some of the data sets are old and of rather limited
precision, especially in the range above
1.4 GeV to about 2.2 GeV, a range which is subject to new measurement
at the VEPP 2000 facility at Novosibirsk. Data from different
experiments are combined by standard methods as recommended by the
Particle Data Group (see e.g.~\cite{Eidelman:1995ny}). In recent years
progress has been due to much better \sigh~ determinations at
Novosibirsk (CMD2, SND)~\cite{CMD206,SND06} and more recently by the
novel radiative return high accuracy measurements by
KLOE~\cite{KLOE08,KLOE10} and \babar~\cite{BABAR09} (see also
~\cite{Davier:2010nc,Hagiwara:2011af}). Typically, vacuum polarization
leads to large corrections and in fact $\alpha(E)$ is steeply
increasing at low $E$ already. So the deviation of $\alpha(m_\mu)$ at
the muon mass scale $m_\mu$ from $\alpha$ gives the big leading
hadronic correction to the muon $g-2$~\cite{FJbook}.  That is why we
need to know the running of $\alpha_\mathrm{QED}$ very precisely at
all scales (see Fig.~\ref{fig:alphaQEDdual}).
Non-perturbative
hadronic effects in electroweak precision observables affect most SM
predictions via non-perturbative effects in parameter shifts, typically:
\be
\sini\:\cosi\: =\frac{\pi \al}{\sqrt{2}\:G_\mu\:M_Z^2} 
\frac{1}{1-\Delta r_i}
\label{eq:cosi}
\ee
where\\[-6mm] 
\ba
\Delta r_i &=&\Delta r_i({\al ,\: \Gmu ,\: M_Z ,}\:m_H,
\:{m_{f\neq t},\:m_t})
\ea
represent the quantum corrections from gauge boson self-energies, 
vertex-- and box--corrections.
Uncertainties obscure in particular the indirect bounds on the Higgs mass
obtained from electroweak precision measurements.
Basic observables like
$M_W$ [$\sinW = 1-M_W^2/M_Z^2$], 
$g_2$ [$\sing = e^2/g_2^2=(\pi \al)/(\sqrt{2}\:G_\mu\:M_W^2)$] or
the vector coupling
$v_f$ [$\sinf = (4|Q_f|)^{-1}\;\left(1-{v_f}/{a_f} \right)\;,\;\;f\neq \nu$]
are related to versions of $\sin^2\Theta_W$ obtained form (\ref{eq:cosi})
and the general form of $\Delta r_i$ reads
\ba
\Delta r_i &=& \dal - f_i(\sini)\:\dro + \Delta r_{i\:\mathrm{remainder}}
\ea
with a universal term $\dal$ which affects the predictions for {$M_W$,
$A_{LR}$, $A^f_{FB}$, $\Gamma_f$,} etc.
Only the $\rho$ parameter in the axial coupling 
$a_f$, which is renormalized by $\rho_f = {1}/{(1-\Delta \rho)}$, is independent
from leading non-perturbative hadronic effects. 

One issue concerning running couplings concerns the question complex
vs. real $\alpha(s)$. For $s\neq0$ (\ref{eq:defdal}) provides the
definition of a complex coupling if we relax from taking the real part
only. A typical example where this matters is the vacuum polarization
correction to be performed on $R(s)$ before it can be used in
(\ref{eq:DI}): $R_{\rm physical} \to R^{(0)}
\doteq (\alpha/\alpha(s))^2\,R_{\rm physical}$. Usually, $\alpha(s)$
is take to be real, i.e., $(\alpha/\alpha(s))^2=|1-\Repa \Pi'(s)|^2$
($\Pi'(0)$ subtracted).  More precisely, one should subtract
$|1-\Pi'(s)|^2=\alpha/|\alpha_c(s)|)^2$ where $\alpha_c(s)$ denotes
the complex
version of running $\alpha$. Typically, corrections from imaginary
parts given by $1- |1-\Pi'(s)|^2/(\alpha/\alpha(s))^2$, are small
$\lsim$ 0.1 \% in non-resonance regions.  However, at resonances
corrections are of order $\sim 1/\Gamma_R$ and thus are large for
narrow resonances.

\section{The coupling $ \alpha_2$, $ M_W$  and $ \sinf$}
Unlike for the electromagnetic coupling, for the $SU(2)_L$ coupling
the hadronic shift cannot be directly obtained by integration of
measured data. There is however a pretty clean way to evaluate
$\Delta^{(5)}_{\mathrm{had}} \alpha_2$, contributing to
\ba
\Delta
\alpha_2 &=&-\frac{e^2}{\sin^2\Theta_W}\, \left[\Repa \Pi'^{3 \gamma}(s)
-\Pi'^{3 \gamma}(0)\right]\epo
\ea
which has been proposed long ago
in~\cite{Jegerlehner:1985gq}. The surprising fact is that the
evaluation of $\alpha_2$ does not require to separate all individual
flavor contributions to recombine them in the proper way. In fact, up
to perturbative or very small contributions the hadronic shift of
$\alpha_2$ is proportional to the self-energy correlation amplitude
$\Pi^{3\gamma}$ where $3$ refers to the 3rd component of the weak
isospin current and $\gamma$ to the electromagnetic current. For the
non-perturbative low energy range, it implies that
the contribution corresponding to the $u$, $d$ and $s$ flavors actually requires no
flavor separation in the SU(3) limit. This makes it possible to calculate $\Delta
\alpha_2$ reliably, because the other heavier 
flavors may be safely separated by relying on pQCD weighting. The
assumption is that the for $N_f>3$ the $N_f-1$ lighter flavors above
the $N_f$ flavor threshold can be evaluated by pQCD. A
detailed discussion of the approximations made is given in Appendix C
of~\cite{Jegerlehner:1985gq}. 
Given ${\Pi}^{\gamma\gamma}_\mathrm{con}={\Pi}^{\gamma\gamma}_{(uds)}+{\Pi}^{\gamma\gamma}_{(c)}
+{\Pi}^{\gamma\gamma}_{(b)}$ for the continuum and ${\Pi}^{\gamma \gamma}_\mathrm{res}\simeq {\Pi}^{\rho}+{\Pi}^{\omega}
+{\Pi}^{\phi}
+{\Pi}^{J/\psi}+{\Pi}^{\Upsilon}$ for the narrow resonances,
we have the relations
\ba
{\Pi}^{3\gamma}_\mathrm{con}\simeq \ha\,{\Pi}^{\gamma\gamma}_{(uds)}+\frac38\,{\Pi}^{\gamma\gamma}_{(c)}
+\frac34\,{\Pi}^{\gamma\gamma}_{(b)}
\ea
for the background contribution and
\ba
{\Pi}^{3\gamma}_\mathrm{res}\simeq \ha\,{\Pi}^{\rho}+\frac34\,{\Pi}^{\phi}
+\frac38\,{\Pi}^{J/\psi}+\frac34\,{\Pi}^{\Upsilon}
\ea
for the resonance contributions. The $\rho-\omega$ mixing contribution
usually included in the ${\Pi}^{\rho}$ taking into account the
isospin $I=0$ component $\omega \to \pi\pi$ in the $\gamma\to \pi\pi
\to \gamma$ channel is to be subtracted via the Gounaris-Sakurai parametrization
(by setting to zero corresponding the mixing parameter).
\begin{figure}
\centering
\includegraphics[height=8.5cm]{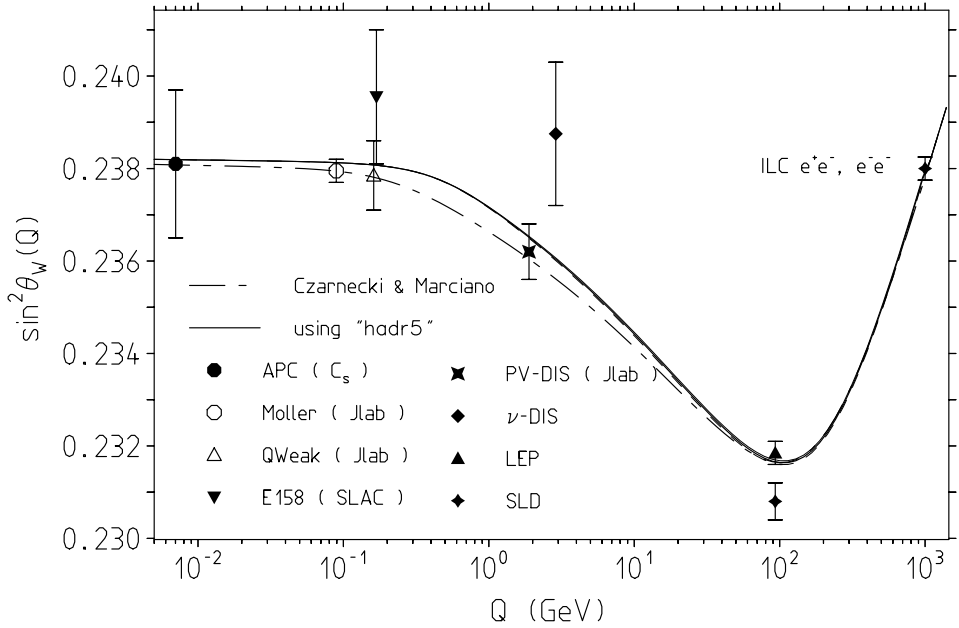}
\caption{$\sin^2 \Theta_W(Q)$ as a function of $Q$ in the space-like
region. Hadronic uncertainties are included but barely
visible. Uncertainties from the input parameter $\sin^2 \theta_W(0)=0.23822(100)$
or $\sin^2 \theta_W(M_Z^2)=0.23156(21)$ are not shown. Future ILC
measurements at 1 TeV would be sensitive to $Z'$, $H^{--}$ etc.}
\label{fig:sin2w} 
\end{figure}
The coupling $ \alpha_2$ can be ``measured'' in a charged current
channel via $M_W$  ($ g \equiv g_2$):
\ba
M_W^2=\frac{g^2\,\varv^2}{4 }=\frac{\pi\, { \alpha_2}}{\sqrt{2}\,G_\mu}
\ea
or via the neutral current channel $\sinf$. In fact here running $
\sinf(E)$ connects the LEP scale mixing parameter to the one of low energy $\nu_e e$
scattering
\ba
\sin^{2}\Theta_{e}(M_Z)=\left\{ \frac{1-\dal_2(M_Z)}{1-\dal(M_Z)}
+\Delta_{\nu_{\mu}e,\mathrm{vertex+box}}
+\Delta \kappa_{e,\mathrm{vertex}} \right\}
\sin^{2}\Theta_{\nu_{\mu}e}\epo
\ea
The first correction from the running coupling ratio is largely
compensated by the \mbo{\nu_\mu} charge radius which dominates the
second term. The ratio
\mbo{\sin^{2}\Theta_{\nu_{\mu}e}/\sin^{2}\Theta_e} is close to 1.002,
independent of top and Higgs mass. Note that errors in the ratio
\mbo{{(1-\dal_2)}/{(1-\dal)}} can be taken to be 100\% correlated and
thus largely cancel.

Above result allow us to calculate non-perturbative hadronic
correction in \mbo{\gamma \gamma}, \mbo{\gamma Z}, \mbo{ZZ} and
\mbo{WW} self energies. Gauge boson self-energies potentially are 
very sensitive to new physics (oblique corrections), which, however,
may be obscured by uncertainties of the non-perturbative hadronic
effects. For complete analytic expressions for electroweak parameter
shifts at one-loop see~\cite{TASI,Jegerlehner:1988ak}. Another
interesting version of running $\sin^2 \Theta_W(Q^2)$ one finds in
\textit{polarized Moeller scattering asymmetries} as advocated by
Czarnecki \& Marciano~\cite{Czarnecki:1995fw} (see
also~\cite{Erler:2004in}). It includes specific bosonic contribution
$\Delta \kappa_b(Q^2)$ such that
\ba
\kappa(s=-Q^2)=\frac{1-\Delta \alpha_2(s)}{1-\Delta \alpha(s)}+\Delta \kappa_b(Q^2)-\Delta \kappa_b(0)
\ea
where\footnote{Here $\Delta \alpha=${\tt dggvap(s,0.d0)} and $\Delta \alpha_2=${\tt degvap(s,0.d0)}
are provided by functions from the package {\tt alphaQED}.} 
, in our low energy scheme, we require $\kappa(Q^2)=1$ at $Q^2=0$.  
Explicitly~\cite{Czarnecki:1995fw},
\ba
\Delta
\kappa_b(Q^2)&=&-\frac{\alpha}{2\pi\,s_W}\,\biggl\{-\frac{42\,c_W+1}{12}\,\ln c_W
+\frac{1}{18}-\left(\frac{r}{2}\,\ln
\xi-1\right)\,\biggl[(7-4z)\,c_W \\ &&+\frac16\,(1+4z)\biggr]
-z\,\biggl[\frac34-z+\left(z-\frac23\right)\,r\,\ln
\xi+z\,(2-z)\,\ln^2\xi\biggr]\biggr\}\cs \nn\\
\Delta
\kappa_b(0)&=&-\frac{\alpha}{2\pi\,s_W}\,\biggl\{-\frac{42\,c_W+1}{12}\,\ln c_W +\frac{1}{18}
+\frac{6\,c_W+7}{18}\biggr\}\cs
\ea
with $z=M_W^2/Q^2$, $r=\sqrt{1+4z}$, $\xi=\frac{r+1}{r-1}$, $s_W=
\sin^2\Theta_W$ and $c_W=\cos^2\Theta_W$. Results
obtained in~\cite{Czarnecki:1995fw} based on one-loop perturbation
theory using light quark masses $m_u=m_d=m_s=100~\mv$ are compared with
results obtained in our non-perturbative approach
in Fig.~\ref{fig:sin2w}.

\section{Adler function controlled split data vs pQCD}
A strategy to exploit the rather precise perturbative QCD predictions
in a optimal well controlled way is to monitor QCD predictions via the Adler function
$D(Q^2)$ in the Euclidean region by comparing theory and data there.
\ba
D(-s) &\doteq&
\frac{3\pi}{\alpha}\:s\:\frac{d}{ds} \Delta \alpha_{\mathrm{had}}(s)
= -\left( 12 \pi^2 \right)\:s\: \frac{d\Pi'_{\gamma}(s)}{ds}\cs\\
D(Q^2)& =& Q^2\;\int\limits_{4 m_\pi^2}^{\infty}ds\,
\frac{R(s)}{\left( s+Q^2 \right)^2} \epo
\ea
Low energies, resonances and thresholds prevent us from making
reliable and precise predictions of $R(s)$ in pQCD. Locally deviations
between data and $R$-predictions can be huge. In contrast, the smooth
function $D(Q^2)$ is easy to compare and deviations show up at low
energies only. A detailed inspection of the time-like approach shows
that pQCD works well in ``perturbative windows'' like 3.00 GeV - 3.73
GeV, 5.00 GeV - 10.52 GeV and 11.50 GeV - $\infty$.  In the space-like
approach pQCD works well for $\sqrt{Q^2=-q^2} > 2.5$
GeV~\cite{Eidelman:1998vc,Jegerlehner:2008rs}. 
Theory is based on results by Chetyrkin, K\"uhn et 
al.~\cite{Chetyrkin:1996cf,Chetyrkin:1997qi}. One thus requires data to
calculate
\be
\Delta \alpha_{\mathrm{had}}(-s_0) = \frac{\alpha}{3\pi} \int_0^{s_0}
dQ^{'2} \frac{D(Q^{'2})}{Q^{'2}}
\ee
up to $s_0=(2.5~\gv)^2$. Equivalently, $\Delta
\alpha_{\mathrm{had}}(-s_0)$ can be directly calculated by
(\ref{eq:DI}) and used in (\ref{eq:alphaAdler}). One 
obtains~\cite{Eidelman:1998vc,Jegerlehner:2008rs}
\ba \bary{lccc}
\Delta\alpha^{(5)}_{\rm had}(-s_0)^{\mathrm{data}} &=& 0.007337 \pm
0.000090 &\\
\Delta\alpha^{(5)}_{\rm had}(-M_Z^2) &=&  0.027460 \pm 0.000134\nn &\\
\Delta\alpha^{(5)}_{\rm had}(M_Z^2) &=&  0.027498 \pm 0.000135 &\epo
\eary
\ea         
The result includes a shift $+0.000008$ from the 5-loop contribution.
The error $\pm 0.000103$ in the perturbative part is added in quadrature.
QCD parameters used are $\alpha_s(M_Z)=0.1189(20)$,
$m_c(m_c)=1.286(13)~[M_c=1.666(17)]~\gv\,,$
and $m_b(m_c)=4.164(25)~[M_b=4.800(29)]~\gv\,$ based on a complete 3--loop
massive QCD analysis~\cite{Kuhn:2007vp}. The latter results are in
agreement with results from lattice
QCD~\cite{Rolf:2002gu,Heitger:2009qf,DellaMorte:2006cb}.
Results based on the Adler controlled split are 
$\Delta \al _{\rm hadrons}^{(5)}(\mz) =0.027498 \pm 0.000135~[0.027510 \pm
0.000218]$ or $\alpha^{-1}(\mz)=128.962 \pm 0.018~[128.961 \pm 0.030]$ 
in braces for comparison the results obtained by the standard approach.

A comparison of error profiles between $\dahz$, $\dah0$ and $a_\mu$
may be found in~\cite{Jegerlehner:2008rs}.  Note that our approach,
with a conservative cut of $\sqrt{s_0}=2.5~\gv$, does not rely
substantially more on pQCD than standard analyses by Davier, H\"ocker
et al.~\cite{Davier:2010nc} and others (see Tab.~\ref{tab:pQCDpart}).
\begin{table}
  \caption{How much pQCD? $\dahz \power{4}$ pQCD part only.}
  \label{tab:pQCDpart}
\begin{tabular}{lcrcr}
\hline
Method & range [GeV] & pQCD && \\
\hline
Standard approach: & 5.2 - 9.5  & 33.50(0.02) && \\
My choice & 13.0 - $\infty$ & 115.69(0.04) &$\to$& {149.19 (0.06)}\\
\hline 
Standard approach: & 2.0 - 9.5  & 72.09(0.07) && \\
Davier et al. & 11.5 - $\infty$ & 123.24(0.05) &$\to$& {195.33 (0.12)}\\ 
\hline
Adler function controlled: & 5.2 - 9.5 & 3.92(0.00) && \\
& 13.0 - $\infty$ & 1.09(0.00) &&\\ 
& $-\infty $~~-~~$ -2.5 $ & 201.23(1.03) && \\
& $-M_Z \to M_Z$ &0.38(0.00) &$\to$& {206.62 (1.03)}\\ 
\hline
\end{tabular}
\end{table}
Further progress is possible due to progress in methods to include the
hadronic $\tau$--decay data~\cite{Jegerlehner:2011ti,Benayoun:2011mm}.

\section{The FORTRAN package {\tt alphaQED}}
\begin{figure}
\centering
\includegraphics[height=12cm]{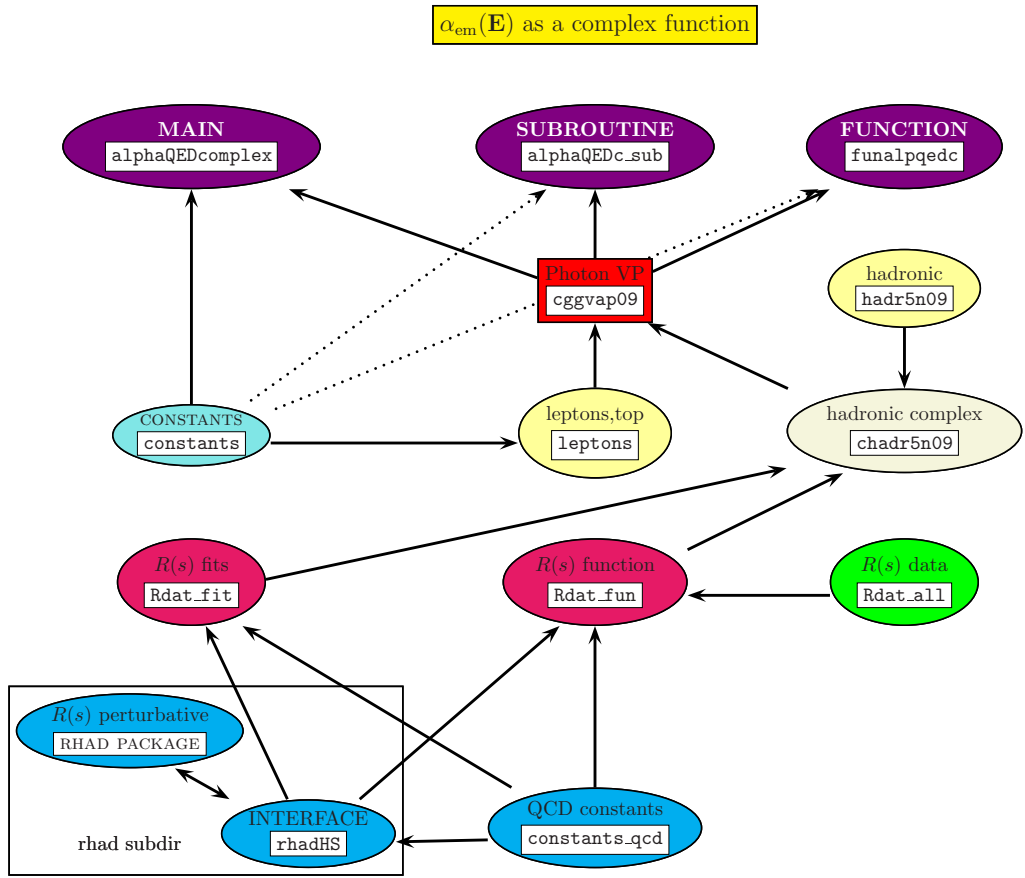}
\caption{Structure of {\tt alphaQEDcomplex}. The corresponding diagram
for {\tt alphaQEDreal} is much simpler as it involves the upper part only.}
\label{fig:alphaQEDcomplex} 
\end{figure}
The FORTRAN package {\tt alphaQED.tar.gz}~\cite{alphaQED}  
for calculating the SM effective 
couplings includes two versions: 
\begin{itemize}
\item
{alphaQEDreal} [FUNCTION {funalpqed}] providing the real
part of the subtracted photon vacuum polarization including hadronic,
leptonic and top quark contributions as well as the weak part
(relevant at ILC energies). Hadronic, leptonic, top and weak
contributions are accessible separately via common blocks 
\begin{verbatim}
      common /resu/dalept,dahadr,daltop,Dalphaweak1MSb
      common /resg/dglept,dghadr,dgetop,Dalpha2weak1MSb
\end{verbatim}
\item
{alphaQEDcomplex} [FUNCTION {funalpqedc}]
provides in addition the corresponding imaginary parts.
See Fig.~\ref{fig:alphaQEDcomplex}. 
\item
corresponding options {alpha2SMreal} and {alpha2SMcomplex} are
available for the {$SU(2)_L$} coupling {$\alpha_2=g^2/4\pi$}.
\end{itemize}
The functions are available for the space-like and the time-like
region.  The complex versions require to install the {\tt Rhad}
package of Harlander and Steinhauser~\cite{Harlander:2002ur} (FORTRAN
package version rhad-1.01 (March 2009 issue)). The latter also
provides the QCD coupling $\alpha_3(s)=\alpha_s(s)$. The imaginary
part given by the bare $R^{(0)}(s)$ is provided in parametrized form
by Chebyshev polynomial fits.  For sample plots I refer to the package
description on my web page
\url{http://www-com.physik.hu-berlin.de/~fjeger/}. The ``organigram'' of
the program is shown in Fig.~\ref{fig:alphaQEDcomplex}.

\acknowledgments

I thank the organizers of the LC 2010 Workshop at Frascati for the
kind invitation, the kind hospitality and for the support.  This work
was supported in part by the European Commission's TARI program under contract
RII3-CT-2004-506078.


\begin{thebibliography}{0}
\bibitem{alphaQED} 
F. Jegerlehner, \textit{The effective fine structure constant
and other SM running couplings}, January 2010, {\tt alphaQED} package download
\url{http://www-com.physik.hu-berlin.de/~fjeger/alphaQED.tar.gz}; see also
  F.~Jegerlehner,
  Nucl.\ Phys.\ Proc.\ Suppl.\  {\bf 162} (2006) 22;
  \textit{The Effective fine structure constant at TESLA energies},
   in *2nd ECFA/DESY Study 1998-2001* 1851-1871; 
  arXiv:hep-ph/0105283.
%
\bibitem{Harlander:2002ur}
R.~V.~Harlander, M.~Steinhauser,
Comput.\ Phys.\ Commun.\  {\bf 153} (2003) 244.
%
\bibitem{Eidelman:1995ny}
  S.~Eidelman, F.~Jegerlehner,
  Z.\ Phys.\  C {\bf 67} (1995) 585.
%
\bibitem{CMD206}
V.~M.~Aulchenko et al.  [CMD-2 Collaboration],
JETP Lett.\  {\bf 82} (2005) 743
[Pisma Zh.\ Eksp.\ Teor.\ Fiz.\  {\bf 82} (2005) 841];
R.~R.~Akhmetshin et al.,
JETP Lett.\  {\bf 84} (2006) 413
[Pisma Zh.\ Eksp.\ Teor.\ Fiz.\  {\bf 84} (2006) 491];
Phys.\ Lett.\  B {\bf 648} (2007) 28.
%
%
\bibitem{SND06}
M.~N.~Achasov et al. [SND Collaboration],
J.\ Exp.\ Theor.\ Phys.\  {\bf 103} (2006) 380
[Zh.\ Eksp.\ Teor.\ Fiz.\  {\bf 130} (2006) 437].
%
\bibitem{KLOE08}
  F.~Ambrosino {\it et al.}  [KLOE Collaboration],
  Phys.\ Lett.\  B {\bf 670} (2009) 285.
%
\bibitem{KLOE10}
  F.~Ambrosino {\it et al.}  [KLOE Collaboration],
  Phys.\ Lett.\  B {\bf 700} (2011) 102.

\bibitem{BABAR09}
  B.~Aubert et al.  [BABAR Collaboration],
  Phys.\ Rev.\ Lett.\  {\bf 103} (2009) 231801.
%
\bibitem{Davier:2010nc}
  M.~Davier, A.~H\"ocker, B.~Malaescu, Z.~Zhang,
  arXiv:1010.4180 [hep-ph].
%
\bibitem{Hagiwara:2011af}
  K.~Hagiwara, R.~Liao, A.~D.~Martin, D.~Nomura, T.~Teubner,
  arXiv:1105.3149 [hep-ph].
%
\bibitem{FJbook}
F.~Jegerlehner, \textit{The Anomalous Magnetic Moment of the Muon},
STMP 226, (Springer, Berlin Heidelberg 2008) Sec. 2.7;
  F.~Jegerlehner, A.~Nyffeler,
  Phys.\ Rept.\  {\bf 477} (2009) 1.
%
%
\bibitem{Jegerlehner:1985gq}
  F.~Jegerlehner,
  Z.\ Phys.\  C {\bf 32} (1986) 195.

\bibitem{TASI}
 F. Jegerlehner, \textit{Renormalizing the Standard Model}, in
    \textit{Testing the Standard Model},   edited by M.
 Cveti\v{c}, P. Langacker, (World Scientific, Singapore) 1991, p. 476;
 {Prog. Part. Nucl. Phys.} {\bf 27} (1991) 32; 
(see \url{http://www-com.physik.hu-berlin.de/~fjeger/books.html}).

\bibitem{Jegerlehner:1988ak}
  F.~Jegerlehner, \textit{Precision Tests of Electroweak-Interaction
  Parameters}, in \textit{Testing the Standard Model}, edited by
  M. Zra\l ek and R. Ma\'nka, (World Scientific, Singapore) 1988, pp. 33-108. 

\bibitem{Czarnecki:1995fw}
  A.~Czarnecki, W.~J.~Marciano,
  Phys.\ Rev.\  D {\bf 53} (1996) 1066;
%
  Int.\ J.\ Mod.\ Phys.\  A {\bf 15} (2000) 2365.
%
\bibitem{Erler:2004in}
  J.~Erler, M.~J.~Ramsey-Musolf,
  Phys.\ Rev.\  D {\bf 72} (2005) 073003.
%
\bibitem{Eidelman:1998vc}
  S.~Eidelman, F.~Jegerlehner, A.~L.~Kataev, O.~Veretin,
  Phys.\ Lett.\  B {\bf 454} (1999) 369;
  F.~Jegerlehner,
\textit{Hadronic effects in $(g-2)_\mu$ and $\alpha_{\rm QED}(M_Z)$: Status and perspectives},
 in \textit{Radiative Corrections},
 edited by J. Sol\`a, (World Scientific, Singapore) 1999, pp. 75--89.
%
\bibitem{Jegerlehner:2008rs}
  F.~Jegerlehner,
  Nucl.\ Phys.\ Proc.\ Suppl.\  {\bf 181-182} (2008) 135
  [arXiv:0807.4206 [hep-ph]].

\bibitem{Chetyrkin:1996cf}
  K.~G.~Chetyrkin, J.~H.~K\"uhn, M.~Steinhauser,
  Nucl.\ Phys.\  B {\bf 482} (1996) 213;
  Nucl.\ Phys.\  B {\bf 505} (1997) 40.
%
\bibitem{Chetyrkin:1997qi}
  K.~G.~Chetyrkin, R.~Harlander, J.~H.~K\"uhn, M.~Steinhauser,
  Nucl.\ Phys.\  B {\bf 503} (1997) 339.
%
\bibitem{Kuhn:2007vp}
  J.~H.~K\"uhn, M.~Steinhauser, C.~Sturm,
  Nucl.\ Phys.\  B {\bf 778} (2007) 192.
%
\bibitem{Rolf:2002gu}
  J.~Rolf, S.~Sint  [ALPHA Collaboration],
  JHEP {\bf 0212} (2002) 007.
%
\bibitem{Heitger:2009qf}
  J.~Heitger,
  Nucl.\ Phys.\ Proc.\ Suppl.\  {\bf 181+182} (2008) 156.
%
\bibitem{DellaMorte:2006cb}
  M.~Della Morte, N.~Garron, M.~Papinutto, R.~Sommer,
  JHEP {\bf 0701} (2007) 007.
%
\bibitem{Jegerlehner:2011ti}
  F.~Jegerlehner, R.~Szafron,
  Eur.\ Phys.\ J.\  C {\bf 71} (2011) 1632.
%
\bibitem{Benayoun:2011mm}
  M.~Benayoun, P.~David, L.~DelBuono, F.~Jegerlehner,
  arXiv:1106.1315 [hep-ph].

\end{thebibliography}
\end{document}